\documentclass[showpacs,twocolumn,prb]{revtex4}
\usepackage{amssymb}
\usepackage{graphicx}

\begin{document}

\title{Theoretical Prediction of Topological Insulators in Thallium-based III-V-VI$_2$ Ternary Chalcogenides
}
\author{Binghai Yan$^1$, Chao-Xing Liu$^{2}$, Hai-Jun Zhang$^{3,4}$,  Chi-Yung Yam$^1$,
Xiao-Liang Qi$^{4,5}$, Thomas Frauenheim$^1$ and Shou-Cheng
Zhang$^4$}
\affiliation{$^{1}$Bremen Center for Computational Materials Science,
Universit$\ddot{a}$t Bremen, Am Fallturm 1, 28359 Bremen, Germany\\
$^{2}$Physikalisches Institut (EP3) and
  Institute for Theoretical Physics and Astrophysics,
  University of W$\ddot{u}$rzburg, 97074 W$\ddot{u}$rzburg, Germany\\
$^{3}$Beijing National Laboratory for Condensed Matter
  Physics, and Institute of Physics, Chinese Academy of Sciences,
  Beijing 100190, China\\
  $^{4}$Department of Physics, McCullough Building, Stanford University,
    Stanford, CA 94305-4045\\
    $^{5}$Microsoft Research, Station Q, Elings Hall, University of
California, Santa Barbara, CA 93106, USA
}
\date{\today}

\begin{abstract}
We predict a new class of three dimensional topological
insulators in thallium-based III-V-VI$_2$ ternary chalcogenides,
including TlBiQ$_2$ and TlSbQ$_2$ (Q = Te, Se and S). These
topological insulators have robust and simple surface states
consisting of a single Dirac cone at the $\Gamma$ point. The
mechanism for topological insulating behavior is elucidated using
both first principle calculations and effective field theory models.
Remarkably, one topological insulator in this class, TlBiTe$_2$ is
also a superconductor when doped with $p$-type carriers. We discuss
the possibility that this material could be a topological
superconductor. Another material TlSbS$_2$ is on the border between
topological insulator and trivial insulator phases, in which a
topological phase transition can be driven by pressure. 
\end{abstract}

\pacs{}
\maketitle

Topological insulators have attracted great attention in condensed matter
physics\cite{qi2010}. Since the first theoretical
prediction\cite{bernevig2006d} and the subsequent experimental
observation\cite{koenig2007}in HgTe quantum wells, several other topological
insulators in three dimensional (3D) bulk materials have been theoretically
predicted and experimentally
observed\cite{fu2007a,hsieh2008,zhang2009,xia2009,chen2009}. In particular,
tetradymite semiconductors Bi$_2$Te$_3$, Bi$_2$Se$_3$, and Sb$_2$Te$_3$ are
predicted to be topological insulators with a large bulk band gap whose surface
state consists of a single Dirac cone\cite{zhang2009}. The mechanism for the
topological insulating behavior in these 3D materials is the band inversion at
the $\Gamma$ point caused by large spin-orbit coupling, similar to the
mechanism first discovered in HgTe quantum wells\cite{bernevig2006d}. The
tetradymite semiconductors have a layered structure consisting of stacking
quintuple layers, making surface preparation particularly simple.

In this work we predict a new class of 3D topological insulators in
the thallium-based III-V-VI$_2$ ternary chalcogenides. These
inversion symmetric topological insulators have a bulk energy gap
and topologically protected surface states consisting of a single
Dirac cone. Unlike the tetradymite semiconductors, these materials
are intrinsically 3D, and do not have a weakly coupled layer
structure. Nonetheless, effective field theory model describing the
band electrons close to the Fermi energy takes the same form as the
model proposed earlier for the tetradymite
semiconductors\cite{zhang2009}, and the mechanism for topological
insulating behavior can be understood in a similar way.

The discovery of topological insulators also inspired the intense search for topological
superconductors \cite{qi2009a,qi2009,schnyder2008,kitaev2009,roy2008}. Time reversal invariant
topological superconductors have a full pairing gap in the bulk and topologically
protected surface states consisting of Majorana fermions, see Fig 1 of Ref.\cite{qi2009a}.
Wheareas Dirac fermions have both particle and hole types, Majorana fermions are
their own anti-particles.\cite{wilczek2009} In the simplest version, the surface state of a 3D
topological superconductor consists of a single Majorana cone, thus containing half
the degree of freedom of the Dirac surface states of a simple 3D
topological insulator. This fractionalization of the degree of freedom introduces
quantum non-locality and is central to the program of topological quantum computing
based on Majorana fermions\cite{nayak2008}. Superfluid He3B phase has been proposed as an candidate
for the 3D topological superfluid state, however, no example of
a topological superconductor state has been found so far.

The most striking property of this new class of topological insulators is the
superconductivity observed in $p$ doped TlBiTe$_2$\cite{Hein1970}.
In this paper, we propose that the $p$ doped TlBiTe$_2$ could inherit the topological
properties from its parent topological insulator, and outline pairing scenarios
under which the topological superconductor state could be realized.

\begin{figure}
    \begin{center}
      \includegraphics[width=3in]{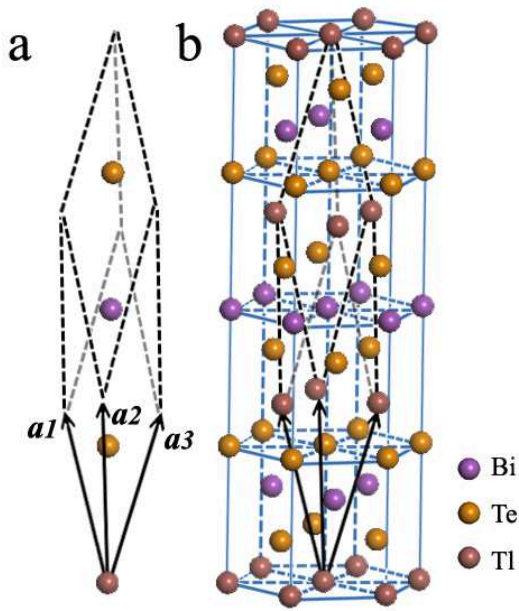}
    \end{center}
\caption{ Crystal structure. (a) Crystal structure of TlBiTe$_2$ with three
primitive lattice vectors denoted as \textit{\textbf{a}}1;2;3, including four
atoms. (b) The equivalent hexagonal lattice of TlBiTe$_2$ with atomic
layers stacked in the sequence -Tl-Te-Bi-Te-. }
    \label{fig:crystal}
\end{figure}

Thallium-based III-V-VI$_2$ ternary chalcogenides have rhombohedral crystal structure
with the space group $D^5_{3d}$ (R$\bar{3}$m), which is similar to tetradymite semiconductors.
We take TlBiTe$_2$ as an example. The crystal structure of TlBiTe$_2$ can be viewed as
the distorted NaCl structure with four atoms in the primitive unitcell, which is shown
in Fig 1 (a). These atoms are placed in layers normal to the
three-fold axis with the sequence -Tl-Te-Bi-Te-\cite{hockings:a03074}.
For Bi$_2$Se$_3$, five atomic layers form a quintuple layer
and the coupling between two quintuple layers is very weak, of the van der Waals type\cite{zhang2009}.
However here the outmost electron shells for Tl, Bi and Te are all p-orbitals ($6p^1$ for Tl, $6p^3$ for Bi
and $5p^4$ for Te) and each Tl (Bi) layer is sandwiched by two Te layers, therefore there is
strong coupling between every two atomic layers for TlBiTe$_2$ and the crystal structure is
essentially 3D. Figure 1(b) shows the 3D layered structure with
the rhombohedral unit cell of TlBiTe$_2$. Similar to Bi$_2$Se$_3$, there is inversion symmetry
for this type of materials and both Tl and Bi sites act as the
inversion center under an inversion operation.

\begin{figure}
   \begin{center}
      \includegraphics[width=3.5in]{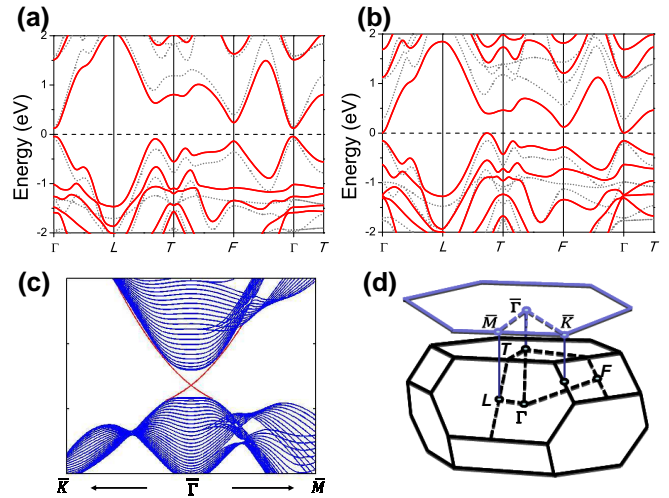}
    \end{center}
    \caption{ Band structure, surface state and Brillouin zone. Band
structure for (a) TlBiSe$_2$ and (b) TlBiTe$_2$ without (gray dotted lines)
and with (red solid lines) SOC. The dashed line indicates the Fermi level.
(c) The surface states (red line) shows a single Dirac cone for TlBiSe$_2$.
The black line shows the band edge for conduction band and valence band.
The dispersion of surface states are calculated from the effective Hamiltonian
(\ref{eq:Hsur}).
(d) Brillouin zone for this class of materials.
The four inequivalent time-reversal-invariant points are $\Gamma$(0;0;0),
$L(\pi;0;0), F(\pi;\pi;0)$ and $T(\pi;\pi;\pi)$. We further project the
3D bulk Brillouin zone into the plane of the atomic layer to obtain
the surface Brillioun zone. }
    \label{fig:band1}
\end{figure}

To investigate the electronic band structure and topological property of these
materials, we performed \textit{ab initio} calculations within the density
functional theory using the Perdew- Burke-Ernzerhof type generalized gradient
approximation\cite{perdew1996}and the projected augmented wave
method\cite{kresee1999} implemented in the \textit{Vienna ab initio simulation
package}\cite{kresse1996}. The plane wave basis is used with energy cutoff of
300 eV. Spin-orbit coupling (SOC) is included except in ionic optimization. We
optimized of lattice parameters and ionic positions first, and then used the
relaxed structure to calculate electronic properties. The optimized lattice
parameters agree with reported
experimental\cite{hockings:a03074,madelung2004,ozer1996} and
theoretical\cite{Hoang2008} results. Our calculated band structures for these
six materials are also consistent with previous calculations\cite{Hoang2008}.
The electronic band structures of TlBiSe$_2$ and TlBiTe$_2$ without (dashed
line) and with SOC (solid line) are shown in Fig. 2 (a) and (b), respectively.
There is a large energy shift for the conduction and valence band and the band
gap shrinks considerably when the SOC is turned on, indicating that the SOC
plays an important role for this type of materials. The conduction band minimum
(CBM) lies at $\Gamma$ point for both TlBiSe$_2$ and TlBiTe$_2$, while the
valence band maximum (VBM) is at $\Gamma$ point for TlBiSe$_2$ and along the
L-T line for TlBiTe$_2$. Therefore TlBiSe$_2$ is a direct gap semiconductor
while TlBiTe$_2$ is an indirect narrow gap semiconductor, consistent with
previous experiments and
calculations\cite{Paraskevopoulos1987,jensen1972,Hoang2008}. In order to
investigate the topological property of the system, we analyze the charge
density of the VBM and CBM at the $\Gamma$ point. Before SOC is included, the
VBM is composed mainly by the Te-$p$ orbital and CBM mainly by the Bi-$p$ and
Te-$sp$ orbitals. After SOC is turned on, VBM and CBM exchange their charge
density characteristics, indicating that a band inversion may occur, just as in
the case of HgTe quantum wells and Bi$_2$Se$_3$ type of materials. To further
determine the topological nature of the system, we follow the parity criteria
proposed by Fu and Kane\cite{fu2007a}, and calculate the product of the parity
eigenvalues of the Bloch wavefunction for the occupied bands at all
time-reversal-invariant momenta $\Gamma, L, F, T $ in the Brillouin zone, both
with and without SOC. For TlSbS$_2$, the product of the parity eigenvalues
remains the same when the SOC is included, while for all the other materials in
this class, the parity eigenvalue of one occupied band (VBM) at $\Gamma$ point
changes after turning on SOC, and the parity eigenvalues for all the occupied
bands at $L, F, T$ do not change. The parity eigenvalues of the total 10
valence bands and the first conduction band at $\Gamma$ point, as well as the
band gaps are listed in Table 1. From the parity analysis we find that
TlSbS$_2$ is a trivial insulator and all the other materials in this class are
strong topological insulators. Although TlSbS$_2$ is a trivial insulator, it
sensitively depends on the lattice constant. We find that about 2\% compressive
strain along the $c$-axis in the hexagonal lattice will induce a band inversion
between the conduction and valence bands and the corresponding pressure is less
than 2 GPa, which is reachable under ambient condition. Therefore this material
opens the exciting possibility to systematically study the topological quantum
phase transition between 3D topological insulators and normal insulators by
tuning pressure.

\begin{figure}
    \begin{center}
    \includegraphics[width=3.5in]{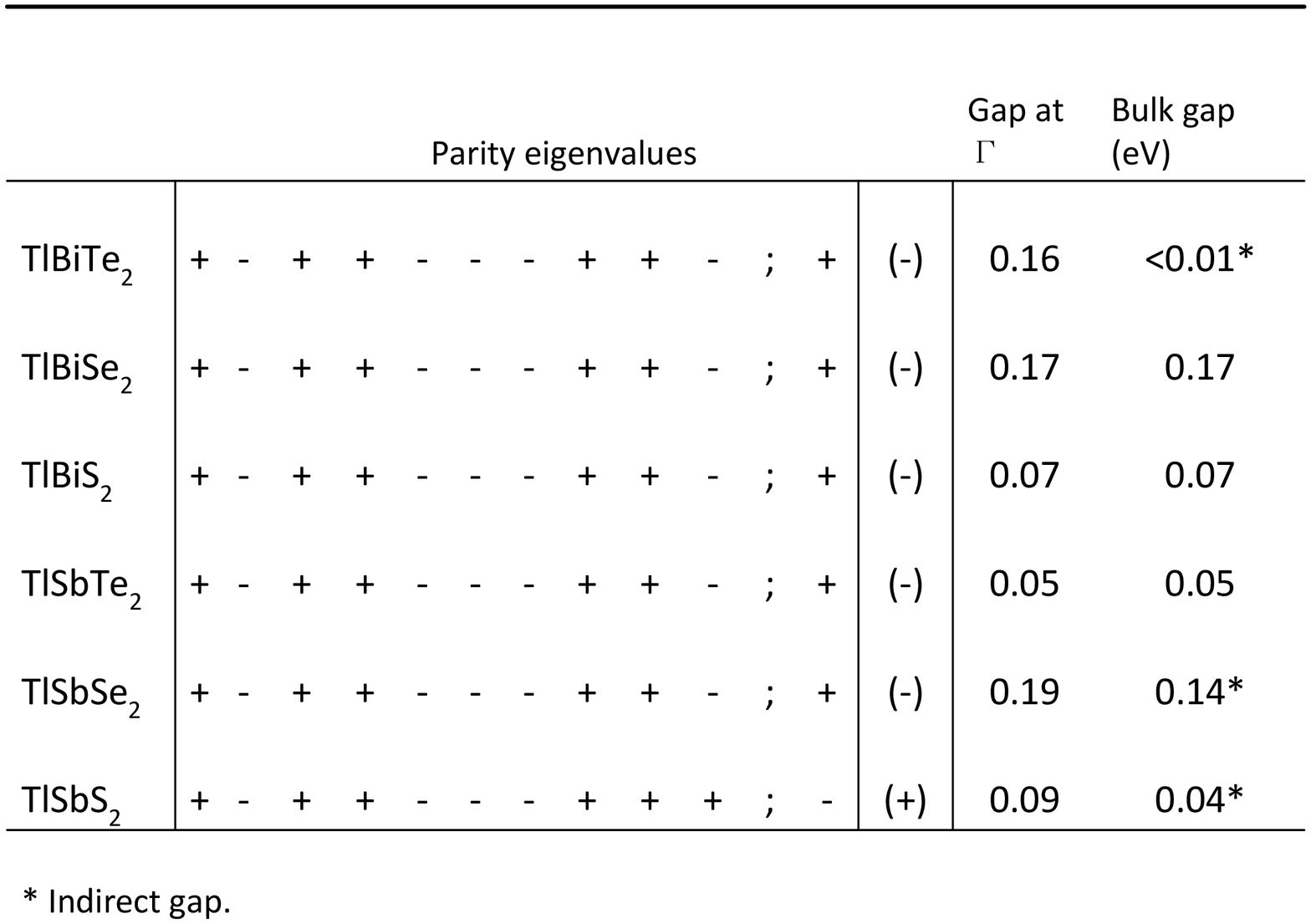}
    \end{center}
    \caption{ Table: the parity of the band at the $\Gamma$ point and the
band gap for the six materials. Here, we show the parity eigenvalues of ten occupied
bands and the lowest unoccupied band. The product of the parity eigenvalues for the ten
occupied bands is given in brackets on the right of each row. We list the band
gap both at the $\Gamma$ point and in the whole Brillouin zone (bulk gap) in
unit of eV.}\label{fig:parity}
\end{figure}


As discussed above, the band inversion only occurs at the $\Gamma$
point, therefore it is helpful to investigate the effective
Hamiltonian near the $\Gamma$ point, which can be constructed from
the symmetry property of the system. To derive the effective
Hamiltonian, we need to identify at $\Gamma$ point the
representations of the crystal symmetry group for both the
conduction and the valence bands. Here we denote the conduction band
as $P1^+$ band and the valence band as $P2^-$ band, where $\pm$
denotes the parity of the corresponding bands. With the spin
degeneracy, four bands ($\left(\left|P1^+,\uparrow\right\rangle,
\left|P2^-,\uparrow\right\rangle,
\left|P1^+,\downarrow\right\rangle,
\left|P2^-,\downarrow\right\rangle\right)$) need to be taken into
account for the effective Hamiltonian. The crystal structure of
TlBiTe$_2$ belongs to $D^5_{3d}$, which is same to that of
Bi$_2$Se$_3$. Thus the wave functions at $\Gamma$ point can also be
classified according to the symmetry group of $D^5_{3d}$, similarly
as we have done for Bi$_2$Se$_3$\cite{liu2010}. According to the
irreducible representation of the space group $D^5_{3d}$, we find
that the conduction band belongs to the $\Gamma_6^+$ representation
while the valence band belongs to the $\Gamma_6^-$ representation.
Consequently, the four band effective Hamiltonian of Bi$_2$Se$_3$ is
still valid here, which in the basis of
$\left(\left|P1^+,\uparrow\right\rangle,
\left|P2^-,\uparrow\right\rangle,
\left|P1^+,\downarrow\right\rangle,
\left|P2^-,\downarrow\right\rangle\right)$
reads\cite{zhang2009,liu2010}
\begin{eqnarray}
    &&H({\bf k})=H_0({\bf k})+H_3({\bf k})\label{eq:Heff}\\
    &&H_0({\bf k})=\epsilon_0({\bf k}){\rm I}_{4\times 4}+\nonumber\\
    &&\left(
    \begin{array}{cccc}
        \mathcal{M}({\bf k})&-iA_1k_z&0&iA_2k_-\\
        iA_1k_z&-\mathcal{M}({\bf k})&iA_2k_-&0\\
        0&-iA_2k_+&\mathcal{M}({\bf k})&-iA_1k_z\\
        -iA_2k_+&0&iA_1k_z&-\mathcal{M}({\bf k})
    \end{array}
    \right)\label{eq:Heff1}
\\
    &&H_3({\bf k})=\frac{R_1}{2}
    \left(
    \begin{array}{cccc}
        0&K_+&0&0\\
        K_+&0&0&0\\
        0&0&0&-K_+\\
        0&0&-K_+&0
    \end{array}
    \right)\nonumber\\
    &&+\frac{R_2}{2}\left(
    \begin{array}{cccc}
        0&-K_-&0&0\\
        K_-&0&0&0\\
        0&0&0&-K_-\\
        0&0&K_-&0
    \end{array}
    \right)
    \label{eq:Heff2}
\end{eqnarray}
with $k_\pm=k_x\pm ik_y$, $\epsilon_0({\bf
k})=C+D_1k_z^2+D_2k_\perp^2$, $\mathcal{M}({\bf
k})=M-B_1k_z^2-B_2k_\perp^2$ and $K_\pm=k_+^3\pm k^3_-$. Here
$H_0({\bf k})$ is the effective Hamiltonian expanded up to the $k$
quadratic term, which preserve the in-plane rotation symmetry.
$H_3({\bf k})$ is the $k$-cubic term which breaks the in-plane
rotation symmetry down to the three fold rotation symmetry.
Combining the ${\bf k\cdot p}$ perturbation theory with the {\it ab
initio} calculation\cite{liu2010}, we can numerically calculate the
parameters of our model, giving $C=-0.045eV$, $M=0.087eV$,
$A_1=1.330eV\cdot$\AA, $A_2=2.821eV\cdot$\AA,
$D_1=6.338eV\cdot$\AA$^2$, $D_2=11.140eV\cdot$\AA$^2$,
$B_1=0.342eV\cdot$\AA$^2$, $B_2=18.225eV\cdot$\AA$^2$,
$R_1=14.367eV\cdot$ \AA$^3$ and $R_2=43.331eV\cdot$\AA$^3$ for
TlBiTe$_2$. The parameters $M>0$, $B_1>0$ and $B_2>0$ indicate that
the system stays in the inverted regime and is topologically
non-trivial. The topologically non-trivial surface states can be
directly calculated from the above four band model by imposing
proper boundary condition\cite{koenig2008,zhang2009,shan2010}.
Taking a semi-infinite sample where the Hamiltonian (\ref{eq:Heff})
applies only for $z>0$, it can be shown that two localized states
$|\psi_\uparrow\rangle$ and $|\psi_\downarrow\rangle$ appear at
$k_x=k_y=0$ which are time-reversal partner of each other. We can
further project the effective Hamiltonian (\ref{eq:Heff}) onto the
subspace spanned by these two localized states, which
yields\cite{liu2010}
\begin{eqnarray}
    H_{sur}=\tilde{C}+\tilde{D}_2k_\parallel^2+\tilde{A}_2(k_x\sigma_y-k_y\sigma_x)
    +\frac{\tilde{R}_1}{2}(k^3_++k^3_-)\sigma_z
    \label{eq:Hsur}
\end{eqnarray}
up to $k^3$. Here the parameters $\tilde{C}$, $\tilde{D}_2$,
$\tilde{A}_2$, and $\tilde{R}_1$ depend on the detail of the
boundary condition\cite{liu2010} and material
parameters\cite{shan2010}. For simplicity, we just take bulk values
for these parameters, the energy dispersion for the topologically
nontrivial surface states are plotted in the Fig 2(c), which shows a
single Dirac cone at $\Gamma$ point, similar to the case of
Bi$_2$Se$_3$. For Bi$_2$Se$_3$, due to the weak coupling between two
quintuple layers, the clean surface can be easily obtained in
experiment. For TlBiSe$_2$, as we have pointed out above, every two
atomic layers are strongly coupled, and consequently, trivial
surface states with dangling bonds may exist on a cleaved surface.
However these trivial surface states will not change the topological
property of the system.


\begin{figure}
    \begin{center}
    \includegraphics[width=3.5in]{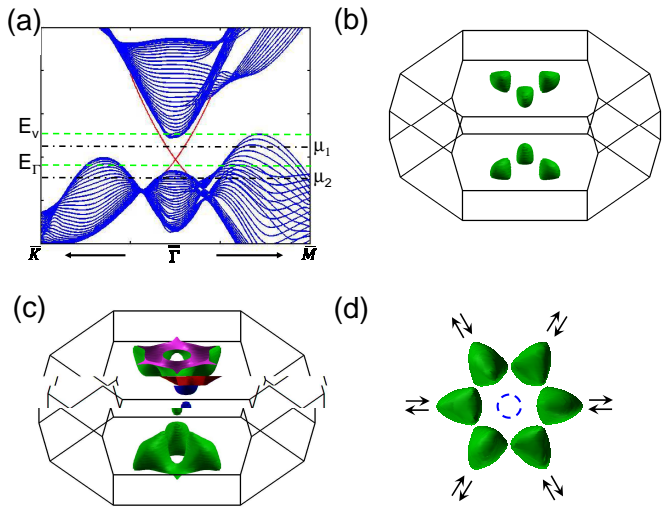}
    \end{center}
    \caption{ Superconductivity in $TlBiTe_2$. a. Bulk and surface states of $TlBiTe_2$. $E_v$ and $E_\Gamma$ are the valence band top in the whole Brillouin zone and at
    $\Gamma$ point, respectively. $\mu_1$ and $\mu_2$ labels two typical positions of chemical potential. b and c. Fermi surface shape corresponding to $\mu_1$ and
    $\mu_2$ respectively. d. Schematic picture of the He$^3$-B type triplet pairing with spin direction correlated with momentum.}\label{fig:sc}
\end{figure}

Among the predicted materials, TlBiTe$_2$ is observed to be a
superconductor with $p$-type carrier.\cite{Hein1970} TlBiTe$_2$ has
an indirect gap, with 6 hole pockets around the $T$ point upon hole
doping, as shown in Fig. 4 b and c. Compared with the other five
materials in this family, which either have an indirect gap
comparable with the $\Gamma$ point gap or have a direct gap, it is
natural to expect that the 6 hole pockets around the $T$ point are
responsible for the superconductivity in TlBiTe$_2$. Depending on
the position of the chemical potential, there are two distinct
superconducting phases possible, as shown in Fig. 4 a. First, when
the chemical potential lies below the top of the valence band $E_v$,
but still above the top of the valence band at the $\Gamma$ point
$E_\Gamma$, topological surface states around the $\Gamma$ point
coexist with the 6 bulk hole fermi pockets. In the superconducting
state, the surface states can also become superconducting due to
proximity effect with the bulk states. As proposed by Fu and
Kane\cite{fu2008}, each vortex of such a superconductor has a
Majorana zero mode, making this system a new candidate for
topological quantum computation\cite{nayak2008}. Compared to the Cu
doped Bi$_2$Se$_3$ superconductor discovered
recently\cite{wray2009}, the coexistence between topological surface
states and bulk superconductivity in TlBiTe$_2$ is much better
defined, because of the well-separation of the surface and bulk
states in momentum space. Second, when the chemical potential lies
below $E_\Gamma$, the topological surface states are not
well-defined any more, and a bulk hole pocket appears around
$\Gamma$ point. When the hole pocket around $\Gamma$ point is small
compared to the pre-existing 6 hole pockets, the pairing symmetry at
the $\Gamma$ point pocket is determined by that of the 6 pockets
through proximity effect in momentum space. If the 6 hole pockets
have the same sign of pairing amplitude, the resulting
superconducting state is topologically trivial. However, since Tl
d-orbital characters are generally present for the wave functions of
the hole pockets, Coulomb correlation effects maybe important, and
inter-pocket repulsive scattering generally prefer opposite signs of
pairing amplitudes on different pockets. This mechanism implies a
negative ``Josephson coupling" between different pockets. However,
due to the three-fold symmetry of the fermi surface, such a coupling
is frustrated, so that the pairing order parameter in the ground
state may become complex. A natural choice of such a complex orbital
pairing symmetry without breaking the time reversal symmetry is a
triplet pairing symmetry similar to the BW state in He3 B-phase, as
illustrated in Fig. 4 d. When the hole pocket around Gamma point
appears, the proximity effect from such a pairing symmetry leads to
a topological superconductor with a nodeless bulk gap, and a gapless
surface state consisting of a single branch of Majorana fermions.


\end{document}